\documentclass[aps,prl,preprint,groupedaddress,showpacs,showkeys]{revtex4-1}
\usepackage{color}
\usepackage{graphicx}
\usepackage{amsmath,amssymb}
\usepackage{amsfonts}

\newcommand{\wz}{{\rm cm}\mbox{$^{-1}$}~}

\newcommand{\chem}[1]{\mbox{{$\rm #1$}}}

\newcommand{\ds}{\displaystyle}

\newcommand{\A}{{\rm {\AA}}}

\newcommand{\VEC}[1]{\mbox{\boldmath $#1$\unboldmath}}

\newcommand{\ket}[1]{\mbox{$\lvert{#1}\rangle$}}
\newcommand{\bra}[1]{\mbox{$\langle{#1}\rvert$}}

\newcommand{\proA}{\mbox{\rm {\A}}\mbox{$^{-1}$}~}

\def\<={\raisebox{-0.5ex}{\small $\stackrel{<}{\sim}$}}
\def\>={\raisebox{-0.5ex}{\small $\stackrel{>}{\sim}$}}
\newcommand{\Eq}[1]{Eq.~({\ref{#1}})}
\newcommand{\Fig}[1]{Figure~{\ref{#1}}}
\newcommand{\Tab}[1]{Table~{\ref{#1}}}

\newcommand{\symgroup}[2]{\mbox{{\boldmath $#1$\unboldmath}$_{\rm #2}$}}

\newcommand{\subi}[2]{\mbox{$ #1_{\rm #2}$}}

\newcommand{\BLAT}[1]{\textcolor{black}{#1}}
\newcommand{\hcwz}{\mbox{$hc\,\wz$}}

\newcommand{\iu}{\chem{i}}

\newcommand{\Exp}[1]{{\rm e}^{\ds #1}}

\newcommand{\Sqe}{S(\VEC{q},E)}
\newcommand{\Sqz}{S(\VEC{q},0)}
\newcommand{\qv}{\VEC{q}}
\newcommand{\GI}{\subi{\Gamma}{i}}
\newcommand{\TI}{\subi{\tau}{i}}

\newcommand{\AD}{\subi{\alpha}{d}}
\begin{document}
\title{Diffusion Rates for Hydrogen on Pd(111)\\ from Molecular
  Quantum Dynamics Calculations} 
\author{Thiago Firmino}
\author{Roberto Marquardt}
\email[corresponding author: ]{roberto.marquardt@unistra.fr}
\affiliation{Laboratoire de Chimie Quantique - Institut de Chimie -
{UMR~7177~CNRS/UdS}\\
Universit\'e de Strasbourg\\
1, rue Blaise Pascal - BP 296/R8 - 
67008 STRASBOURG CEDEX - France}
\author{Fabien Gatti}
\affiliation{CTMM, Institut Charles Gerhardt - 
UMR~5253 CNRS/Universit\'e de Montpellier 2\\
34095 MONTPELLIER Cedex 05 - France}
\author{Wei Dong}
\affiliation{Laboratoire de Chimie - UMR~5182 CNRS/Ecole Normale
Sup{\'e}rieure de 
Lyon\\ 46, All{\'e}e d'Italie,
69364 LYON Cedex 07 - France}
\date{\today}
\begin{abstract}
Diffusion rates are calculated on the basis of 
van Hove's formula for the 
dynamical structure factor (DSF) related to particle scattering 
at mobile adsorbates. 
The formula is evaluated quantum mechanically using eigenfunctions 
obtained from 
three dimensional realistic models for 
H/Pd(111) derived from first principle
calculations. Results are compatible with experimental
data for H/Ru(0001) and H/Pt(111), if one assumes that the total rate
obtained from the DSF is  
the sum of a diffusion and a friction rate. A
simple kinetic model to support this assumption is presented. 
\end{abstract}
\maketitle
The diffusion of adsorbed particles is an important process
intervening in heterogeneous catalysis. Yet, and despite the
significant progress achieved in the past decades in the domain of
surface science, our knowledge about such elementary steps in
catalysis remain modest, both experimentally and
theoretically. In the long time domain of milliseconds to seconds,
scanning tunneling microscopy (STM) is capable of unraveling some of
the details of this motion. For instance, Jewell {\em et. al.} report
on quantum tunneling of isolated hydrogen atoms adsorbed on Cu(111)
terraces, which are observed with STM and a spatial resolution of a
few tenths of a nanometer~\cite{Sykes:2012}. These experiments,
carried out at 5~K, 
show the potential technological application arising from the 
interaction between the mobile adsorbates leading to the formation of 
self-assembled clusters. However, the time resolution of 
these experiments does not allow us to follow the motion of
the H atoms in real time, interpreted in these papers as arising from
tunneling. 

The motion of hydrogen atoms adsorbed on metal surfaces has been
explored with picosecond time resolution in \chem{^3He} spin-echo
experiments for H/Pt(111)~\cite{Ellis:2010} and
H/Ru(0001)~\cite{Ellis:2013}. The primary result from these
experiments is the intermediate scattering function (ISF)
$I(\VEC{q},t)$, where $\VEC{q}$ is the wave vector related to the
momentum transferred from the 
scattered \chem{^3He} atoms to the hydrogen atoms moving on the
surface, and $t$ is the time. In~\cite{Ellis:2013}, experimental
results obtained at several temperatures are rationalized by
path-integral molecular dynamics 
calculations and
quantum transition-state theory. From these analyses, the onset of the
quantum tunneling 
regime was shown to occur at about 70~K, for the H/Ru(0001) system. 

In the experimental work, 
\BLAT{ the diffusion rate of the adsorbed particles is obtained from
  adjustments of time-dependent model 
exponential functions to the ISF and varies as a
function of the momentum transferred to the adsorbed particles. 
The diffusion rate is related to   
  quasi-elastic broadening of the}  
dynamical structure factor (DSF) $\Sqe$, which is the 
the temporal Fourier transform of the ISF. 
The ISF is the spatial Fourier transform of the
pair correlation function proposed by L. van
Hove~\cite{vanHove:1954}, who also derived a general expression for
the DSF in terms of the eigenvalues and eigenfunctions pertaining to
the stationary vibrational states of the
adsorbates. \BLAT{Alternatively, one might then determine the diffusion
  rate by inspection of the full 
width at half 
maximum (FWHM) of the DSF.} 

\BLAT{In the present
work}, we modify van Hove's formula as follows:
\begin{equation}
\label{SqE}
\Sqe =
\sum\limits_n\;P_n\;\sum\limits_m\;\left\lvert\sum\limits_k^N\bra{m}\Exp{\iu\qv\VEC{x}_k}\ket{n}\right\rvert^{\ds
  2}\;L(E;(E_m-E_n),\Gamma_{nm}). 
\end{equation}
In this equation, $\ket{n}$ and $\ket{m}$ are vibrational eigenstates of the
scattering centers at energies $E_n$ and $E_m$; $P_n$ is the Boltzmann
population distribution; $\VEC{x}_k$ is the position vector of the
adsorbed particle $k$ ($k=1,\ldots,N$). $L(E;E_0,\Gamma)$ is a normalized 
Lorentzian distribution peaked at $E_0$ and having a full width at
half maximum (FWHM) $\Gamma$. In the original work of van Hove, this 
formula is given with a $\delta$-function instead of the Lorenztian.  
The energy width 
$\Gamma_{nm}$ can be related to the finite lifetimes of vibrational eigenstates. 
State $\ket{n}$ has the lifetime $\tau_n =
h/(\pi\,\gamma_n)$, where $\gamma_n$ is the 
width (FWHM) of the energy distribution of this state in the set of
the true eigenstates of the full system including electronic motion 
or the motion of substrate atoms (phonons), and $h$ is the Planck
constant. The overall width (FWHM) arising from the combination of states
$\ket{n}$ and $\ket{m}$ in~\Eq{SqE} is $\Gamma_{nm} =
\gamma_n+\gamma_m$.  

\BLAT{In a recent work~\cite{roma:68}, we explored the prospect
  of~\Eq{SqE} to calculate diffusion rates of the adsorbate from a
  purely quantum  
mechanical treatment of the dynamics. In the present work,
we evaluate~\Eq{SqE}} using vibrational eigenvalues and eigenfunctions 
derived from a potential energy surface (PES)
for the H/Pd(111) system~\cite{Dong:2010} \BLAT{and a realistic model
  for the lifetimes of these states.}  
In our three dimensional study, a single H atom is considered ($N =1$,
in~\Eq{SqE}),
mimicking a low coverage degree of the substrate. 
Experimental diffusion rates for this system have not been determined so far.

While vibrational eigenstates can be calculated rather
straightforwardly, the determination of vibrational lifetimes is more
involved and data are hardly found. 
Depopulation of vibrational eigenstates of adsorbates on 
metal substrates via formation of electron-hole pairs is expected to
proceed on the picosecond time scale~\cite{Harris:1992,Wolf:2006b}, or even
faster~\cite{Saalfrank:2007,Tremblay:2013}.   
This is about the time scale that can be reached with the \chem{^3He}
spin-echo technique. In~\cite{Tremblay:2013},  
lifetimes for the lowest excited vibrational states in the most stable
H/Pd(111) adsorption sites have been
calculated to be around 500~fs to 1.5~ps. The corresponding energy
broadening range of 2.6 to 0.9~meV is up to 
three orders of magnitude larger than the broadening due to the
diffusion of the adsorbates typically 
observed in the aforementioned \chem{^3He} spin-echo experiments. The
relaxation rate due to the 
coupling to phonons is probably much smaller~\cite{Tully:1992}. 

For converged numerical evaluations of~\Eq{SqE}, sums over many states are
needed, typically 50 to 200. As we do not know the lifetimes for the
entire set of states, we make an ad hoc model assumption based on the
findings from~\cite{Tremblay:2013}: vibrational ground states,
i.e., node-less states at the stable adsorption sites, have a
lifetime of 1~\chem{\mu s}, i.e. a width of about 1.3~neV ; all
vibrationally excited 
states have an  
intrinsic lifetime of \BLAT{about $\TI = 527$~fs}, which corresponds to an
intrinsic energy broadening $\gamma_{\rm i}= h/(\pi\TI) =
2.5$~meV. In practice, the widths used in the Lorentzians of~\Eq{SqE} will
therefore be 2~neV, nearly 2.5~meV and 5~meV, if 
vibrational ground states occur doubly, simply or not at all in a term
of the sum. 

We use the Multiconfiguration Time Dependent Hartree (MCTDH) program
suite~\cite{mctdh:2012,Meyer:2000} to calculate vibrational eigenstates of the
adsorbed particles. The analytical function derived
in~\cite{Dong:2010} was used to generate a natural potential
representation~\cite{Meyer:1996b,Meyer:1998} with the aid of
the ``potfit'' program 
contained in the MCTDH program package. Eigenfunctions were then
calculated within the MCTDH program with the aid of the \BLAT{``block
relaxation 
method''~\cite{Gatti:2008}}. 

The system is represented in a set of
non-Euclidean coordinates (see~\Fig{grid} below) to fully exploit its 
periodicity. Note that the kinetic energy operator contains
non-diagaonal terms, when such coordinates are
used~\cite{Saalfrank:2009}. 
For this three dimensional study, the analytical PES 
from~\cite{Dong:2010} was first modified such as to describe a single
hydrogen atom adsorbed on the palladium substrate. To achieve this,
all hydrogen-hydrogen two-body terms, as well as the
hydrogen-palladium-hydrogen three-body terms in the PES defined
in~\cite{Dong:2010} were zeroed, and the function definition was
changed such as to depend on the coordinates of a single hydrogen
atom. 
\begin{figure}
\includegraphics[width=16cm]{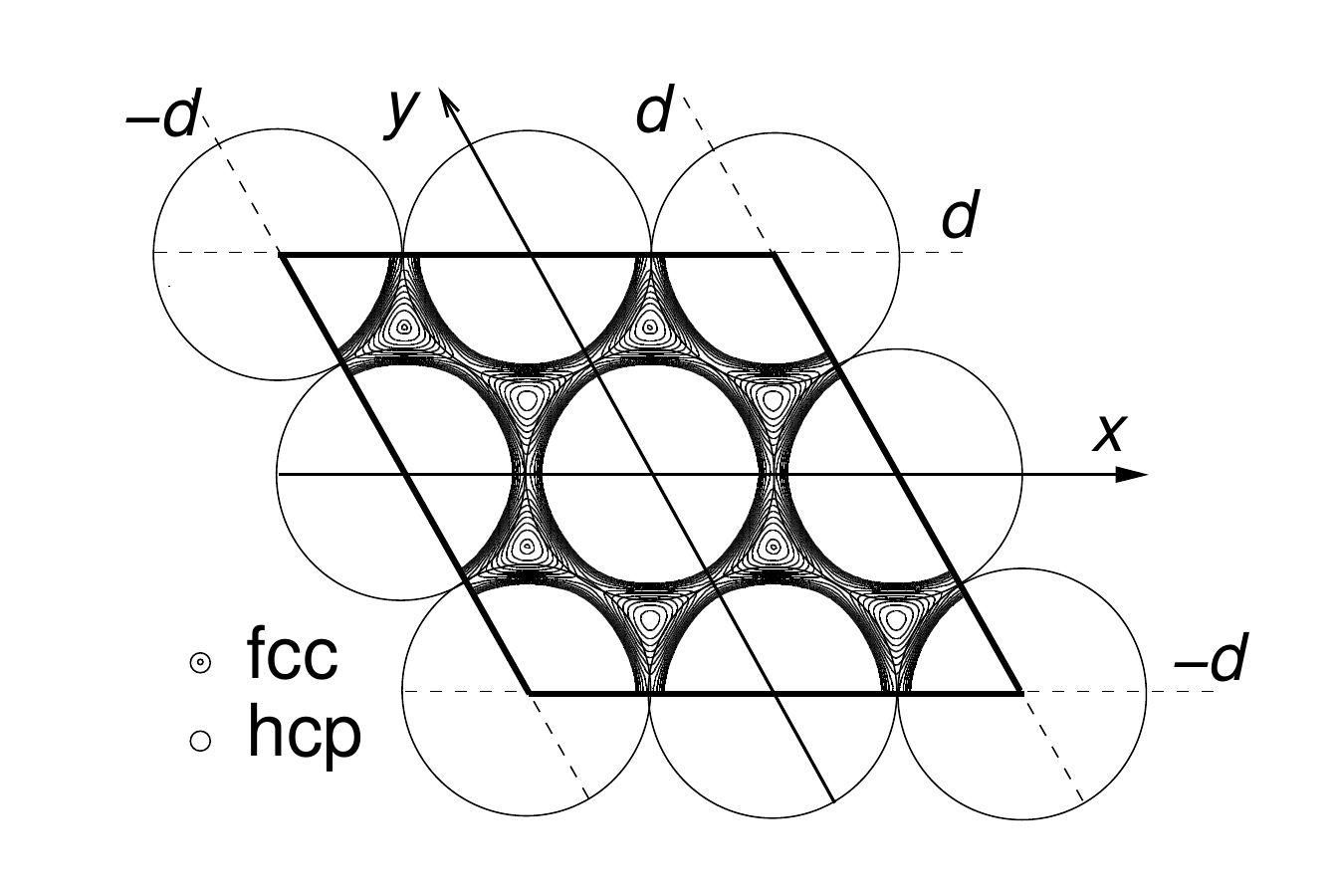}
\caption{\label{grid} Scheme of the $(2\times 2)$ surface cell
  used to 
  characterize the H/Pd(111) system. $x$ and $y$ are the skewed
  coordinates used in the 
  dynamics. Palladium
  atoms are indicated by 
  the large spheres of diameter $d$ ($d = 275.114$~pm is the Pd-Pd
  bulk distance on the PES from~\cite{Dong:2010}). A section of the
  PES for atomic hydrogen at $z = 90$~pm is superimposed on the
  scheme. Contour
  lines are separated by 200~\hcwz, the contour line 
  indicated at most stable fcc site is 50~\hcwz (with a dot at the
  center), the highest, outermost lines shown are at
  2650~\hcwz; lines around hcp sites are at 250~\hcwz.}
\end{figure}
A graphical representation of a
section of the PES is shown in~\Fig{grid}. More details on the
calculations are given in the supplemental material.

This figure also shows the $(2\times 2)$ surface cell underlying the
present calculations. The stable adsorption sites denoted as ``fcc''
and ``hcp'' are clearly indicated. On this
PES, the hcp site is 
about 190~\hcwz less stable than the fcc site and the barrier between
the two sites is at about 1150~\hcwz above the fcc site. 
There are 4 fcc and 4 hcp sites per unit cell. 
The hcp/fcc occupation ratio is
about 0.37 at room temperature, and we can therefore assume that the
occupation of sites is 
approximately homogeneous, which makes the present study mimic a
coverage degree of 12.5\%, for the H/Pd(111) system. 

  Each site has a local
  \symgroup{C}{3v} symmetry, such that per site two vibrational modes
  parallel to 
  the substrate and one mode perpendicular to it can be
  expected. Per site type there will hence be 4 vibrational ground
  states, and 12 vibrationally excited states, namely 8 states of parallel modes  
  and 4 states of perpendicular modes, all arranged in
  levels of quasi-isoenergetic states. Tunneling may split these 
  levels. The present calculations yield that the ground state levels
  at both the ``fcc'' and ``hcp'' sites remain 
  degenerate. However, the 
  vibrationally excited levels split into two blocks for each type of
  mode. The splitting could in principle be observed by high
  resolution spectroscopy. \Tab{transitions} summarizes expected
  transitions, assuming that they occur vertically on top of each
  adsorption site. 

\begin{table}[h]
\caption{\label{transitions} Wavenumbers of the fundamental transitions for
H/Pd(111) 
  in \wz.}
\begin{tabular}{@{}l|c|*2{r}|r|r@{}}
\hline\hline
modes
&
site 
&
\multicolumn{3}{c|}{theory}
&
exp~\cite{Conrad:1986}
\\\cline{3-5}
&
&
\multicolumn{2}{c|}{this work}
&
ref.~\cite{Saalfrank:2009}
&
\\
\hline
parallel      &fcc                 &  743.6~(5) &  744.1~(3) &  717.4   & 774.3   \\
              &hcp                 &  726.8~(3) &  730.8~(5) &          &         \\
\hline                                                      
perpendicular &fcc                 &  1047.6~(1)&  1058.6~(3)&  922.4   & 1016.3  \\ 
              &hcp                 &  1000.2~(3)&  1010.8~(1)&          &         \\
\hline\hline                 
\end{tabular}
\end{table}

Column ``this work'' in~\Tab{transitions} 
  reports four sets of vertical transitions for each mode, two sets of
  transitions per site type. These sets arise
  from the tunneling splitting of degenerate vibrationally excited levels
  and remaining degeneracies are indicated by the numbers in
  parentheses. In refs~\cite{Saalfrank:2009}
  and~\cite{Conrad:1986}, only one 
  value is reported per transition. The present results for
  vibrational wavenumbers are comparable to the previously reported
  values. \Tab{transitions} reports  
  only the fundamental transitions. We note here that the 
  calculated overtone spectrum indicates the presence of strong anharmonic
  resonances, which will have an important influence on the short time
  diffusion dynamics. These results will be
  published elsewhere. 

Vibrational states like those of the H/Pd(111) system in a $(2\times
2)$ surface cell can be cast into a
coarse grained level structure. Levels can be identified by the
vibrational quantum number related to a specific adsorption site, while
nearly degenerate states within a 
level compose a dense structure of states. Similar level structures
arise for larges surface cells. Consequently, the DSF can
be decomposed into a sum $\Sqe = \sum_l\,S_l(\qv,E)$, where $l = 1,
2,\ldots$ denotes a vibrational level. 

\begin{figure}[b]
\includegraphics[width=12cm]{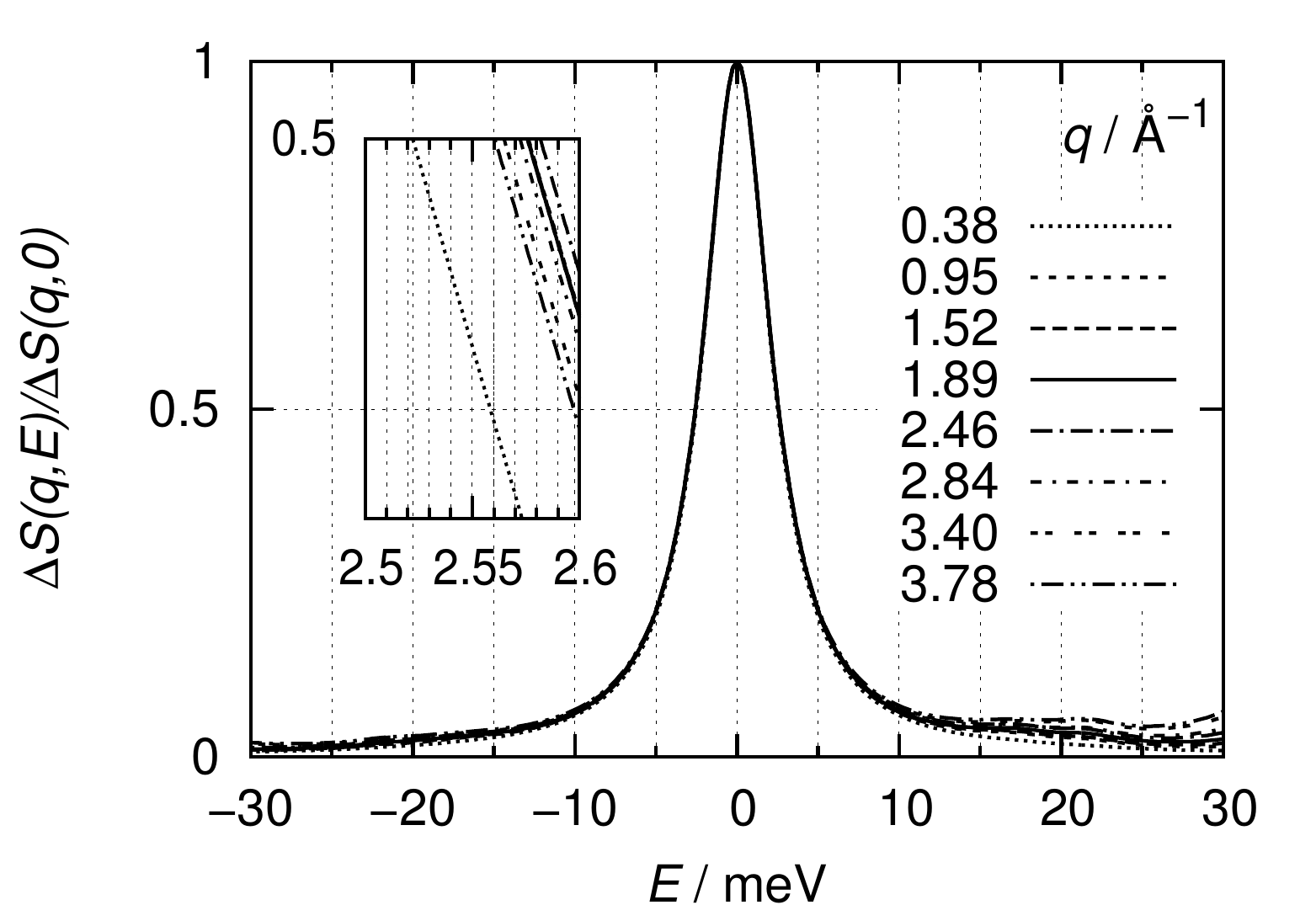}
\caption{\label{HPd} $\Delta S(q,E)/\Delta S(q,0)$ as a function of $E$ and $q$
  for atomic hydrogen on Pd(111). The inset is a magnification of the
  function (right wing). Lines are cubic spline interpolations.}
\end{figure}

Evaluation of~\Eq{SqE} with the eigenvalues and widths discussed above
yields a very 
narrowly peaked function at $E=0$, the \BLAT{width of which} is \BLAT{$2$}~neV. This width 
corresponds to the model lifetime assumed in this work for the
vibrational ground states. It depends only very feebly on $\qv$. Note
that, as the Lorentzians 
in~\Eq{SqE} are energy normalized, the form of the DSF at $E\sim 0$ is
dominated by the contributions from the level of ground states.
It is reasonable to omit these contributions, as they do
apparently not influence the DSF further; in particular, they do not
lead to any diffusion broadening. In the following, we consider
therefore the differential DSF
\begin{equation}
\label{dSqE}
\Delta S(\qv,E) = S(\qv,E) - S_1(\qv,E),
\end{equation} 
where $S_1(\qv,E) \approx S_1(0,E)$ is the contribution to $S(\qv,E)$
from the level of ground states.  
\Fig{HPd} shows the normalized function $\Delta\Sqe/\Delta\Sqz$ along the 
$\langle 1\,1\,\bar{2}\,0\rangle$ direction, for $T = 250$~K. 
For this direction, the scalar
product $\qv\cdot\VEC{x}$ is evaluated by the expression $q\,(x - y/2)$, because of the
non-orthogonality of the $x$ and $y$ axes.

At a first sight, the FWHM of this function depends again 
little on the transferred momentum wave number $q$. It is 
 roughly given by $\GI = 5$~meV, as expected for the intrinsic energy width
adopted for the excited vibrational states in the model calculations
of the present work. 
When magnified 
one sees, however, a neat progression of lines (insert in
\Fig{HPd}). Clearly, it is the {\em differential width} 
$\Delta\Gamma = \Gamma - \GI$ that varies with $q$. 

\begin{figure}[b]
\includegraphics[width=14cm]{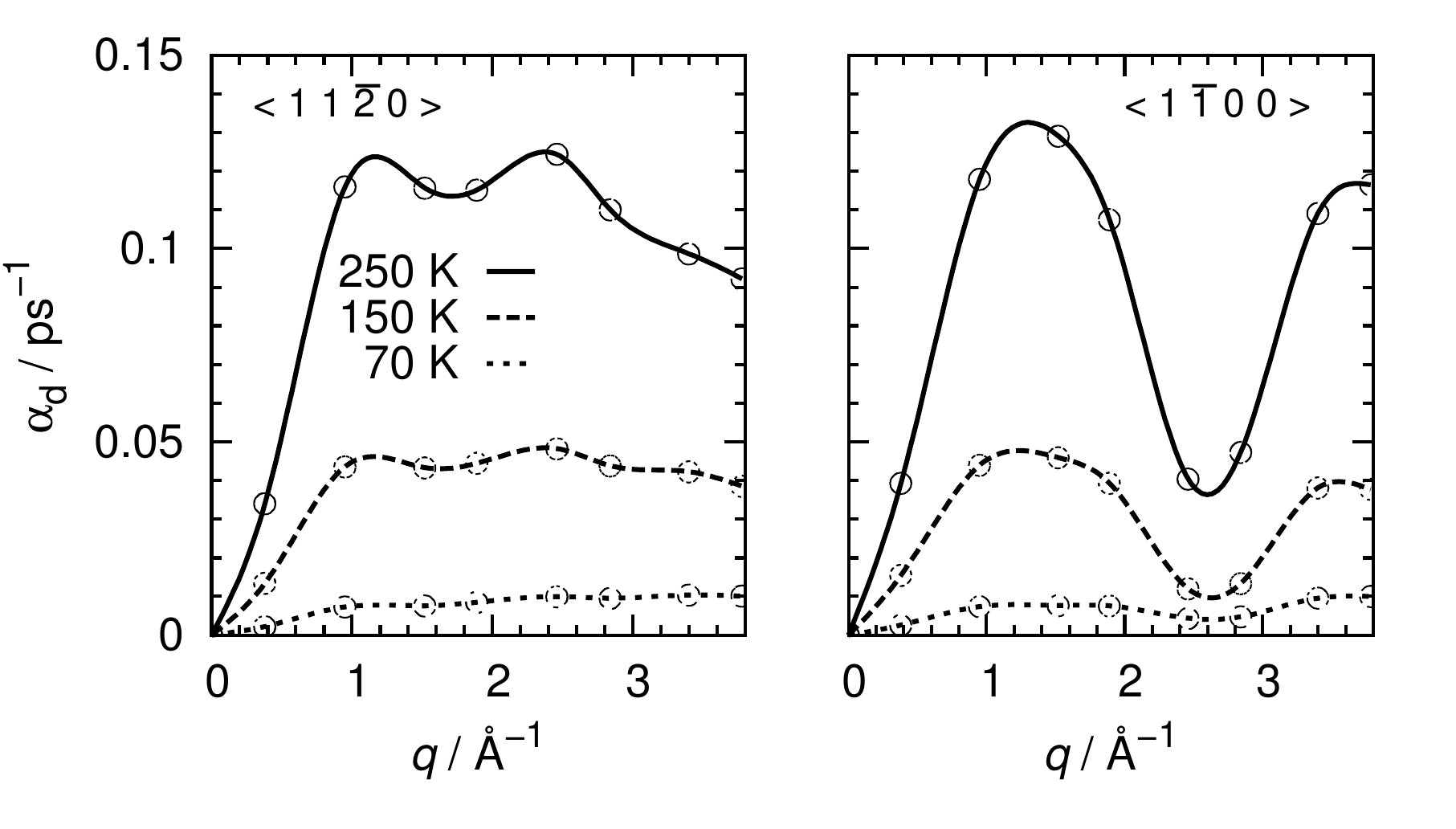}
\caption{\label{Aq} 
Diffusion rate (see
  text) along two crystallographic 
directions and several temperatures, as indicated. Lines are cubic
spline interpolations.
}
\end{figure}

To show this variation in detail, we plot in~\Fig{Aq}  
the rate 
\begin{equation}
\AD =
\pi\,\Delta\Gamma/h
\end{equation} 
$\approx 0.759\,634\;\chem{ps^{-1}}\times
  \Delta\Gamma/\chem{meV}$. The figure shows $\AD$ along the two 
crystallographic directions 
$\langle 1\,1\,\bar{2}\,0\rangle$ 
and 
$\langle 1\,\bar{1}\,0\,0\rangle$ and several temperatures. 
Note that, for the latter crystallographic direction, the scalar
product $\qv\cdot\VEC{x}$ is evaluated by the expression
$q\,\sqrt{3}/2\,(x - y)$. 

The solid line in the figure on the left hand
  side corresponds to the variation obtained in~\Fig{HPd}. 
To our knowledge, there is
currently  no comparable experimental result for the H/Pd(111) system.
We may compare
the present theoretical result, however, with results obtained
in the \chem{^3He} Spin Echo experiments for the diffusion rates 
for the H/Ru(0001)~\cite[figure 1 c]{Ellis:2013} and
H/Pt(111)~\cite[figure 2]{Ellis:2010} systems, at similar coverage
degrees ($\sim 0.1$ mono-layer). We see that $\AD$
reproduces well the 
general behavior of the experimentally obtained function;  
the variation range for $\AD$ is about a factor
10 larger than that observed for H/Ru(0001), 
\BLAT{
and of the same order as 
that for H/Pt(111) ($\langle 1\,1\,\bar{2}\,0\rangle$
direction). Despite the remaining quantitative differences, the
comparison with the experimental results leads} us to call $\AD$ the {\em diffusion
  rate}. 

\begin{figure}
\includegraphics[width=14cm]{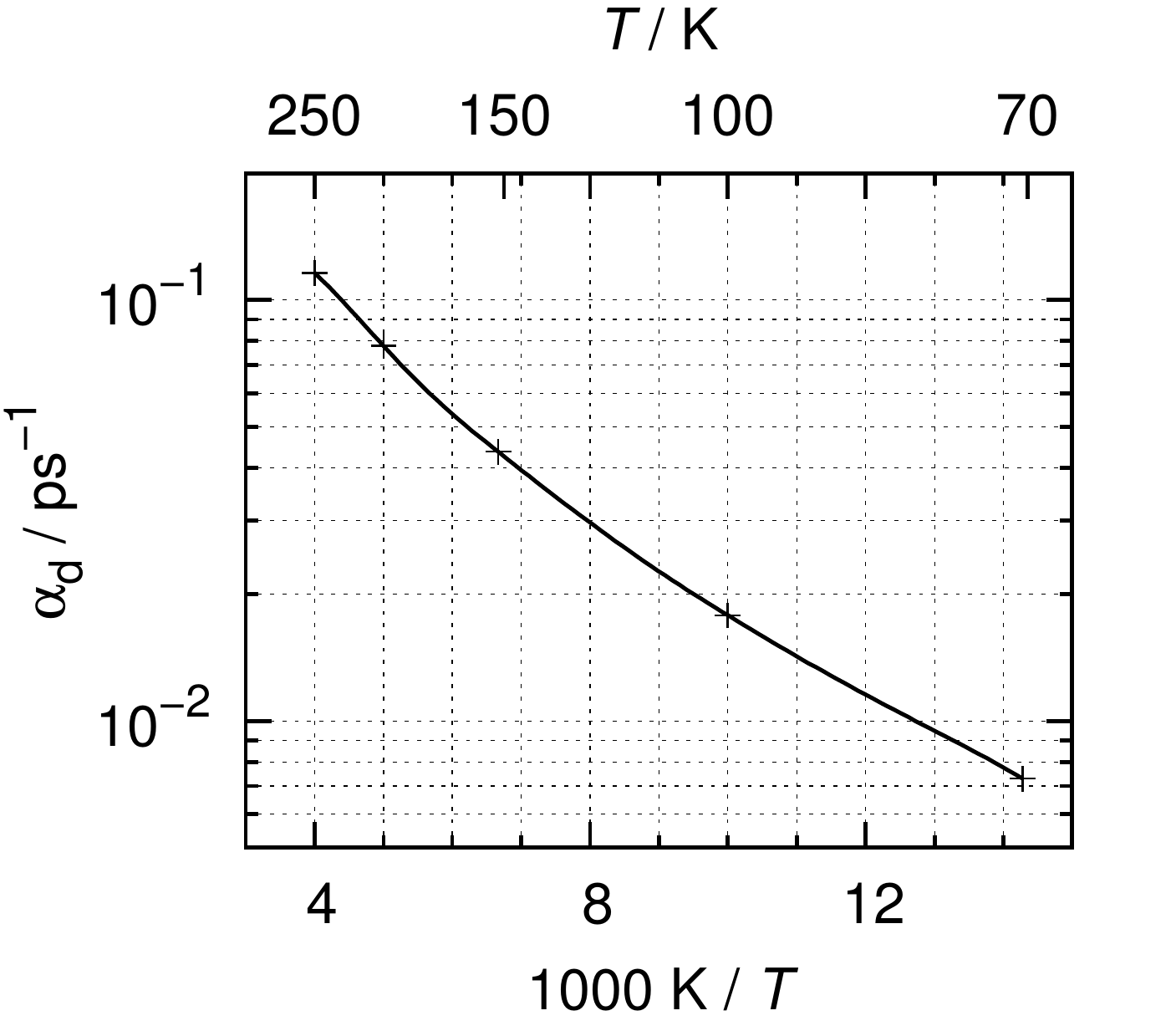}
\caption{\label{AT} 
Arrhenius plot of the temperature dependence of $\AD$; 
\BLAT{$\langle 1\,1\,\bar{2}\,0\rangle$} crystallographic direction and $q =
0.95\;\proA$. 
}
\end{figure}

As an additional \BLAT{support} for our interpretation of $\AD$ as
being the same diffusion rate as that 
determined in the \chem{^3He} Spin Echo experiments, \Fig{AT} shows an
Arrhenius plot of
the temperature dependence of $\AD$. This result is again \BLAT{quite} similar
to that obtained for H/Pt(111)~\cite[figure 3]{Ellis:2010} and
\BLAT{predicts} that also the diffusion rate of hydrogen on palladium
should have a
non classical behavior.  

We can explain that the differential width corresponds indeed to the diffusion
rate. We have developed a theory for the calculation of the ISF from
the vibrational structure of the adsorbates. Variations of the
differential width are shown to be related with the 
vibrational structure of the adsorbates \BLAT{and with the overlap of individual
lines as a function of the intrinsic width}. A detailed account
of the theory as well as a discussion of the quantum effects will be
presented elsewhere. In the present paper we 
develop a simple kinetic model to 
rationalize the variability of the differential width as a function of
the transferred
momentum. Starting from the kinetic model for jump
diffusion~\cite{ChudleyElliot:1961}, we extend it to 
include relaxation by friction. We write the equation of motion for
the probability of finding a 
particle at position $\VEC{x}$ at time $t$ as~\cite{Allison:2010} 
\begin{equation}
\label{model}
\frac{\partial P(\VEC{x},t)}{\partial t} =
\frac{1}{\tau}\;\left(\sum\limits_k\,(P(\VEC{x}+\VEC{y}_k,t) -
P(\VEC{x},t))\right) - \frac{1}{\TI}\,P(\VEC{x},t). 
\end{equation}     
The first part of the right hand side involves summation over lattice
vectors and corresponds to the original jump
diffusion model; it leads to the well known classical exponential
decay for the ISF, if one assumes that the $P(\VEC{x},t)$ equals $G_{\rm
  s}(\VEC{x},t)$, the self-part of the pair correlation function. This
decay can even be a more involved model expression of several exponential
functions~\cite{Allison:2010}, or it can be replaced by an even more
complicated function that one might determine from quantum
dynamics. The second part on the right hand side of~\Eq{model} leads
to a simple exponential decay of the ISF. Now, if the first part leads
to a quasi-exponential decay, as it apparently does, the total decay
rate is then given as the sum of the decay rates of the two
parts. Consequently, it is the 
difference of the total width and the width due to relaxation by
friction that depends strongly on the momentum transfer, and which
is generally related to the diffusion motion.

We have also evaluated~\Eq{SqE} for the six dimensional system
\chem{H_2/Pd(111)}, using the same assumptions regarding the intrinsic
life times of vibrationally excited states. Preliminary results
indicate that the form of 
$\AD$ changes significantly \BLAT{and that comparison with the 
experimental results for hydrogen on platinum becomes more
favorably}. When these results are confirmed, and 
experimental results for the diffusion rate of \BLAT{hydrogen on
  palladium}  
are available, we should 
have an additional source of information regarding the structure of
adsorbed hydrogen on palladium. The computational method described in
this paper can be extended to calculations of the diffusion dynamics
on other transition metal atoms, which 
can then be discussed in relation with the  
findings from ref.~\cite{Sykes:2012}.     

Finally, a comment about the lifetimes of vibrationally excited states
is \BLAT{appropriate}. In the simplified treatment of the present calculations, a
generic realistic value for the life time of excited vibrational
states has been 
adopted. Relevant life times from ref.~\cite{Tremblay:2013} differ only little
from the adopted value and if these are used \BLAT{instead of the generic
one for the few states for
which they are known}, results change insignificantly. However, if the
values for intrinsic life times change \BLAT{by a factor
of 2 or more,} so does the calculated
diffusion rate~\cite{roma:68}. Therefore, the calculation method for the diffusion rates of
adsorbates developed in the present work
is, in combination with their experimental determination, a
new interesting tool to also determine relaxation rates of adsorbates. 
\begin{acknowledgments}
This work was carried out within a research program from the {\em Agence
Nationale de la Recherche} (project ANR 2010 BLAN 720~1). We thank ANR
for the generous financial support, as well as CNRS and Universit{\'e} de
Strasbourg.    
\end{acknowledgments}
\newcommand{\Aa}[0]{Aa}
\end{document}